\documentclass[twocolumn]{aastex63}
\usepackage{amsmath}
\newcommand{\brvs}{Br\"unt-V\"ais\"al\"a}

\submitjournal{ApJ}
\shorttitle{Wave Heating and lithium}
\shortauthors{Jermyn \& Fuller}

\begin{document}

\title{Wave heating during the helium flash and lithium-enhanced clump stars}

\correspondingauthor{Adam S. Jermyn}
\email{adamjermyn@gmail.com}

\author[0000-0001-5048-9973]{Adam S. Jermyn}
\affiliation{Center for Computational Astrophysics, Flatiron Institute, New York, NY 10010, USA}

\author{Jim Fuller}
\affiliation{TAPIR, Mailcode 350-17, California Institute of Technology, Pasadena, CA 91125, USA}

\begin{abstract}
Red Clump stars have been found to be enhanced in lithium relative to stars at the tip of the Red Giant Branch (TRGB), which is unexpected in current stellar models. At the TRGB, stars undergo the helium flash, during which helium burning briefly generates roughly $10^9 \, L_\odot$ of power and drives vigorous convection within the star's core. The helium-burning shell excites large fluxes of internal gravity waves.
Here we investigate whether or not these waves can deposit enough heat to destabilize the hydrogen-burning shell, generate a convection zone there, and thereby drive the Cameron-Fowler process to enhance surface $^{7}\mathrm{Li}$.
We study this with detailed stellar evolution models, and find that while the waves deposit $\sim 10^6 \, L_\odot$ near the hydrogen-burning shell, and while this generally does produce a convection zone, the resulting convection does not reach high enough to merge with the envelope, and so cannot explain enhancements to surface $^{7}\mathrm{Li}$.
\end{abstract}


\section{Introduction}  \label{sec:intro}

Recently,~\citet{2020NatAs.tmp..139K} reported that nearly all Red Clump stars are lithium-rich, with abundances approximately thirty times greater than stars at the tip of the Red Giant Branch (TRGB).
This poses a problem for standard models of low-mass stellar evolution, which predict essentially monotonic lithium depletion with time~\citep{2020A&A...633A..34C}.
At the same time, the timing of the lithium production suggests that the solution to this problem is related to the helium flash, a dramatic stage of stellar evolution which occurs between the TRGB and the Red Clump~\citep{1967ZA.....67..420T} and which may cause a variety of unusual stellar phenomena~\citep{2021arXiv210600582B}.

The typical clump star lithium abundances found in~\citet{2020NatAs.tmp..139K} (see also~\citealt{2021MNRAS.tmp.1175D}) are only $\mathrm{A(Li)}\approx 0.6$, far below the usual lithium rich definition of $\mathrm{A(Li)} \gtrsim 1.5$. Some of these are likely upper limits (Chaname et al. 2021), so many clump stars probably have lower lithium abundances.~\citet{2021yCat..36510084M} show that while roughly one third of low-mass clump stars in clusters are lithium enhanced relative to the upper RGB, the rest of the sample could have much lower lithium abundances. Additionally, while typical upper RGB field stars are dominated by low-mass stars ($M \lesssim 1.3 M_\odot$) with very low lithium abundances, high-mass stars ($M \gtrsim 2 M_\odot$) represent a much larger fraction of clump stars, and these stars suffer far less lithium depletion during the main sequence and RGB, likely  explaining some of the lithium enhanced clump stars found in recent spectroscopic surveys (Chaname et al. 2021). One must also be cautious when comparing observed abundances to models, whose lithium abundances are very sensitive to mass, metallicity, convective mixing length, and thermohaline mixing. Still, it appears that a substantial fraction of low-mass clump stars have higher lithium abundances than their TRGB progenitors, requiring lithium production around the time of helium ignition.

The fraction of very lithium rich red giants that have $\mathrm{A(Li)} > 1.5$ is very small, roughly 1\%~\citep{2021NatAs...5...86Y,2021ApJ...914..116G,2021MNRAS.505.5340M}. The majority of these are clump stars~\citep{2019ApJ...880..125C,2019ApJ...878L..21S}, most of which appear to be near the start of helium burning~\citep{2021ApJ...913L...4S}. These lithium rich giants are more likely to be rapidly rotating (though most still rotate slowly), and nearly all of the super lithium rich giants are low-mass clump stars~\citep{2021MNRAS.505.5340M}. The rarity of these stars points to an uncommon evolutionary channel such as binary interactions~\citep[e.g.][]{2019ApJ...880..125C,2020ApJ...889...33Z}. Since close binary interactions or mergers occur in a few percent of red giants~\citep[e.g.][]{2020ApJ...895....2P}, binary scenarios occur at about the right frequency to match very lithium rich red giant production. However, it is also possible that most clump stars go through a very brief super Li-rich phase soon after the helium flash, which is erased on $\approx \mathrm{Myr}$ time scales~\citep{2021ApJ...913L...4S}

A promising mechanism to produce lithium is chemical mixing followed by the Cameron-Fowler process.
The hydrogen burning shell generates $^{7}\mathrm{Be}$ via the pp-II chain~\citep{1971ApJ...164..111C}.
If $^{7}\mathrm{Be}$ is mixed into the convective envelope it can then undergo electron capture and produce $^{7}\mathrm{Li}$~\citep{2000A&A...358L..49D}.
This suggests that the helium flash somehow causes mixing between the hydrogen burning shell and the envelope~\citet{2011ApJ...730L..12K,2016ApJ...819..135K}.
\citet{Schwab_2020} indeed showed that mixing the hydrogen burning shell with the convective envelope during the helium flash would produce lithium abundances on the Red Clump similar to those observed by~\citet{2020NatAs.tmp..139K}.

One possible way to generate the required mixing is for internal gravity waves emitted by the helium convection zone during the flash to \emph{heat} the hydrogen burning shell, driving convection within the shell and/or extending the convective envelope down to the burning shell.
This could then dredge $^{7}\mathrm{Be}$ into the envelope, which would naturally produce $^{7}\mathrm{Li}$ in the aftermath of the flash.
We have investigated this mechanism quantitatively, and report our findings in this work.

\section{Big Picture} \label{sec:story}

\begin{figure*}
\centering
\includegraphics[width=1\textwidth]{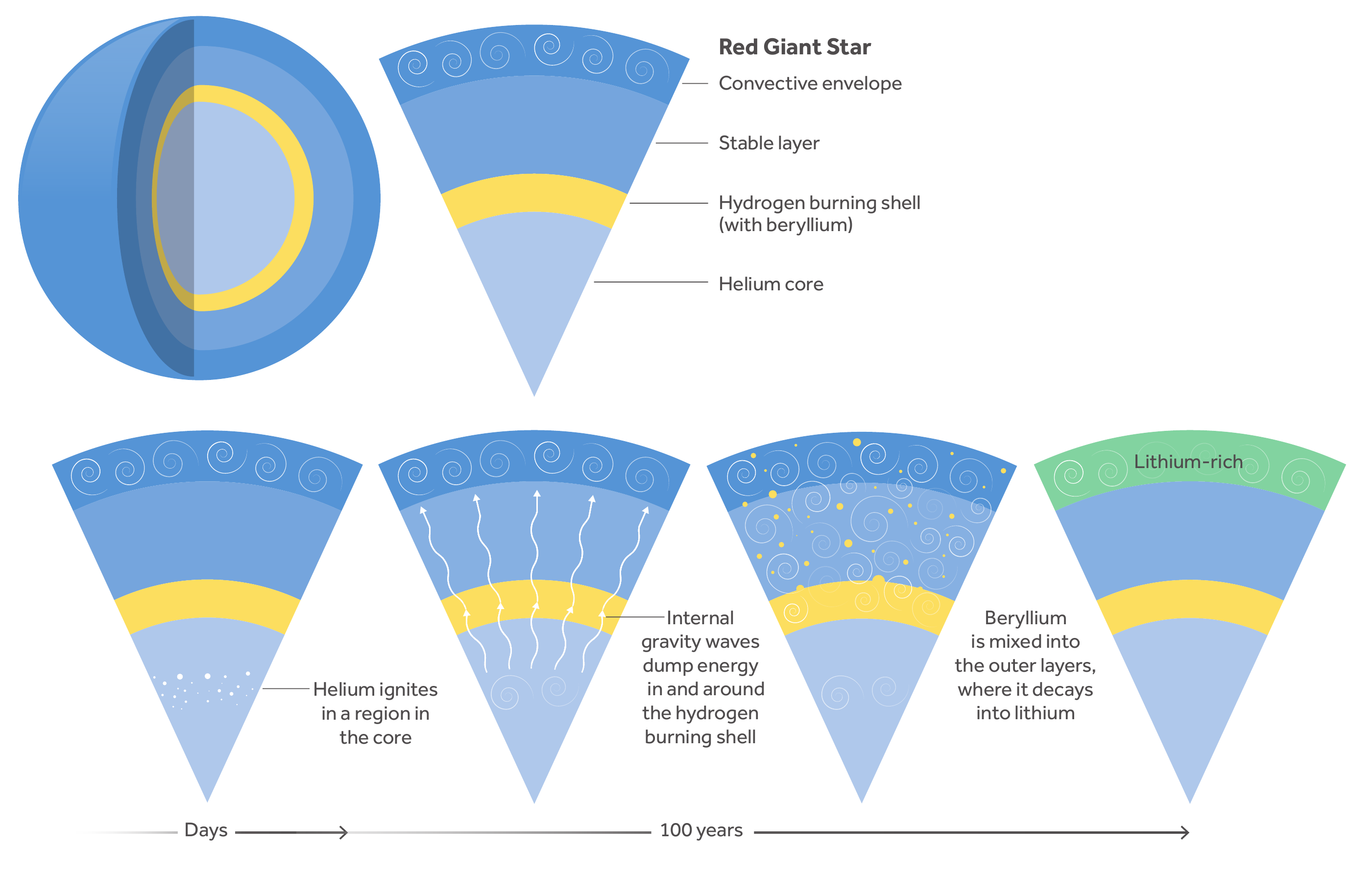}
\caption{(First) The structure of a red giant star at the start of the helium flash is shown schematically. The helium burning shell is convective and excites internal gravity waves which travel into the adjacent radiative zones. The waves that travel upwards first reach the hydrogen burning shell, where $^{7}\mathrm{Be}$ is produced, and then the convective envelope. (Second) Waves damp primarily in the hydrogen burning shell and near the boundary of the convective envelope. (Third) The waves deposit heat, which hypothetically pulls the base of the convective envelope down to the hydrogen burning shell, mixing $^{7}\mathrm{Be}$ into the envelope. (Fourth) After the helium flash the envelope retreats upwards and $^{7}\mathrm{Be}$ undergoes electron capture via the Cameron-Fowler process, forming $^{7}\mathrm{Li}$.}
\label{fig:schema3}
\end{figure*}

The possibility we investigate is shown schematically in Figure~\ref{fig:schema3}.
The helium flash begins with the degenerate ignition of helium in a shell in the core of the star.
Helium burning releases substantial amounts of energy, with a peak luminosity of more than $10^{9} L_\odot$.
This destabilizes the entropy gradient in the burning shell and causes convection, which then carries the most of the nuclear luminosity.
Convection then excites internal gravity waves (IGW) in the adjacent radiative zones.

When the outgoing waves reach the hydrogen burning shell, the steep composition gradient there makes their wavelengths short.
This decreases the radiative damping time, so the waves preferentially damp and deposit their energy in this shell.
Similarly, because the thermal diffusivity rises towards the base of the convective envelope, waves preferentially damp there.
In both cases the wave heat increases the temperature gradient and hence destabilizes the entropy gradient, deepening the convective envelope and generating a new convection zone in the hydrogen burning shell.
Eventually the envelope convection zone extends downward into the hydrogen burning shell, dredging up its burning products, including $^{7}\mathrm{Be}$.

After the peak of the helium flash, the wave heating subsides and the convective envelope retreats upward.
The $^{7}\mathrm{Be}$ then undergoes the Cameron-Fowler process, producing $^{7}\mathrm{Li}$.
Importantly, this occurs once the base of the envelope is cool enough that $^{7}\mathrm{Li}$ is not destroyed by burning at its base.

In what follows we study this picture quantitatively using the Modules for Experiments in Stellar Astrophysics
\citep[MESA][]{Paxton2011, Paxton2013, Paxton2015, Paxton2018, Paxton2019} stellar evolution software instrument.
We find that this basic story does not work for reasonable estimates of the wave amplitudes and frequencies, suggesting that elevated lithium abundances on the Red Clump, if confirmed, likely require an alternative explanation.

\section{Wave Model} \label{sec:model}

We now explain how we compute the wave heating in our stellar models.

\subsection{Wave Excitation} \label{sec:excitation}

Convection excites both pressure waves and internal gravity waves at the radiative-convective boundary.
Because the helium convection zone is subsonic most of the wave power is emitted as internal gravity waves (IGW)~\citep{1990ApJ...363..694G}, so we neglect pressure waves.
\citet{1990ApJ...363..694G} and~\citet{2013MNRAS.430.2363L} each estimated the power and spectrum of these waves.
While they find the same total wave power, the spectra differ because of different assumptions about the magnitude of $k_\perp h$, where $k_\perp$ is the horizontal wave-vector and $h$ is the pressure scale-height.

Because the helium flash takes place near the center of the star, the scale height is of order the radius ($h \approx r$).
The horizontal wave-vector 
\begin{align}
k_\perp \approx \frac{\sqrt{l(l+1)}}{r},
\end{align}
so
\begin{align}
h k_\perp \approx r k_\perp \approx \sqrt{l(l+1)} \ga 1.
\end{align}
As a result the spectrum of~\citet{2013MNRAS.430.2363L} is more appropriate.

A complication in directly using this spectrum is that it was derived on the assumption of an infinite Boussinesq convection zone with uniform properties, whereas in our stellar models the properties of convection change on a length-scale of order the pressure scale-height $h$.
Moreover the spectrum in equation~\eqref{eq:LQ13} reflects a product of the excitation of evanescent waves in the convection zone and a transmission factor controlling their emergence into the adjacent radiative zone.
In cases where properties such as the convective frequency vary spatially these two components must be separated and calculated using the properties where they occur: excitation at each point in the convection zone and emergence just at the boundary.

We address these complications for non-Boussinesq systems with spatially-varying parameters in Appendix~\ref{sec:excitation}, obtaining a wave source
\begin{align}
\frac{dS}{dr d\log \omega d\log k_\perp} \propto (k_\perp h)^5 \left(\frac{\omega}{\omega_c}\right)^{-15/2},
	\label{eq:LQ13}
\end{align}
where $\omega_{\rm c} = \pi v_{\rm c} / 2 h$ is the convective turnover frequency, $v_{\rm c}$ is the convection speed, $h$ is the pressure scale-height, and $S$ is the wave source, which when integrated over the convection zone and multiplied by a transmission factor yields a wave luminosity in the radiative zone.

Equation~\eqref{eq:LQ13} is valid for IGW so long as
\begin{align}
	\omega > \omega_{\rm c} \max\left[1, (h k_\perp)^{2/3}\right],
	\label{eq:boundary_frequency}
\end{align}
For lower frequencies we assume a white noise spectrum for convective kinetic energy in $\omega$.
This produces a spectrum
\begin{align}
	\frac{dS}{d\log \omega} \propto \omega,
\end{align}
scaled so that $dS/d\log \omega$ matches equation~\eqref{eq:LQ13} at the lower frequency boundary (equation~\ref{eq:boundary_frequency}).
That is, if $\omega_{\rm bound} = \omega_{\rm c} \max\left[1, (h k_\perp)^{2/3}\right]$ is the boundary frequency then for $\omega < \omega_{\rm bound}$ we use
\begin{align}
	\frac{dS}{d\log \omega d\log k_\perp} = \left.\frac{dS}{d\log \omega d\log k_\perp}\right|_{\rm bound} \left(\frac{\omega}{\omega_{\rm bound}}\right)^2.
	\label{eq:whitenoise}
\end{align}

We normalize the overall spectrum so that in the spatially-invariant Boussinesq limit the total wave luminosity emitted in each direction by a thick convection zone is
\begin{align}
	L \approx \beta L_{\rm conv} \mathcal{M}_{\rm conv},
	\label{eq:flux_avg}
\end{align}
where $\mathcal{M}_{\rm conv}$ is the Mach number of the convection zone averaged using the evanescent wave eigenfunctions and $\beta$ is a free parameter accounting for the fact that the normalization of the wave luminosity is only known from order-of-magnitude estimates.

\subsection{Wave Propagation}

To calculate wave propagation we begin with the adiabatic dispersion relation~\citep{2002RvMP...74.1073C}
\begin{align}
	c_s^2 k_r^2 + \frac{S_\ell^2}{\omega^2}\left(\omega^2-N^2\right) = \omega^2-\omega_{\rm ac}^2,
	\label{eq:dispersion}
\end{align}
where $c_s$ is the sound speed,
\begin{align}
	S_{\ell}^2 = c_{\rm s}^2 \frac{\ell(\ell+1)}{r^2} = c_s^2 k_\perp^2.
\end{align}
and the acoustic cutoff frequency is given by
\begin{align}
	\omega_{\rm ac}^2 = \frac{c_s^2}{4H^2}\left(1 - 2\frac{dH}{dr}\right)
\end{align}
where $H \equiv |dr/d\ln \rho|$ is the density scale-height.

With equation~\eqref{eq:dispersion} we identify the WKB turning points where $k_r^2 = 0$~\citep{2007AN....328..273G}.
These separate propagating regions from evanescent ones.
Within the WKB approximation we compute the reflection and transmission coefficients across each turning point in each direction and use this to propagate the wave luminosity throughout the model.
In doing so we account for both travelling and standing modes.

\begin{figure}
\centering
\includegraphics[width=0.48\textwidth]{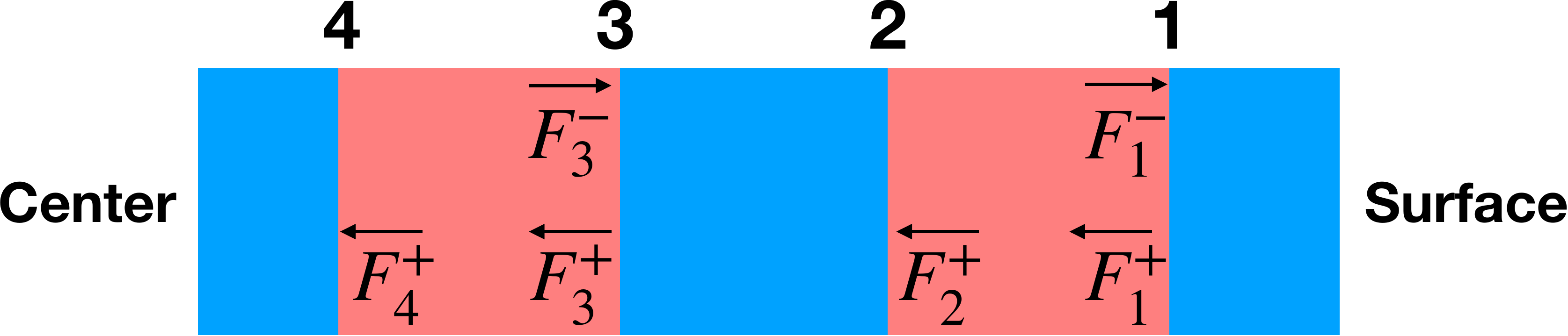}
\caption{A sequence of evanescent (blue) and propagating (red) regions are shown. The luminosity $L_{j}^{-}$ goes towards the surface at turning point $j$ and the luminosity $L_{j}^{+}$ goes towards the center from the same. In both cases the luminosities are defined on the propagating side of the turning point.}
\label{fig:schema5}
\end{figure}

We index the boundaries between these regions by $j=0,1,...,n$, where boundary $0$ is the surface of the star and boundary $n$ is the center.
We refer to the wave luminosity crossing boundary $j$ heading towards the surface by $L_j^-$ and that heaidng towards the center by $L_j^+$.
Because the wave luminosity is only well-defined in propagating regions the luminosity crossing the boundary always refers to the propagating side of the boundary.
Figure~\ref{fig:schema5} shows an example model labelled with these conventions.

We assume that the waves are in steady state, such that the wave power entering a region between boundaries $j$ and $j+1$ ($L_{j}^{+} + L_{j+1}^{-}$) equals the wave power leaving the region ($L_{j+1}^+ + L_{j}^{-}$) plus the wave power lost to radiative damping inside that region ($L_{\rm damp}^{j,j+1}$).
That is,
\begin{align}
	(L_{j}^{+} + L_{j+1}^{-}) - (L_{j+1}^+ + L_{j}^{-}) = L_{\rm damp}^{j,j+1}.
	\label{eq:steady}
\end{align}
Assuming the wave amplitudes are small enough to be linear, this equation factors across frequencies $\omega$ and quantum numbers $\ell,m$ so we may process each $(\omega,\ell,m)$ independently.

We relate the luminosities crossing each boundary to each other using the WKB transmission and reflection coefficients.
For simplicity we compute these coefficients at each boundary in the scenario of a wave approaching from just one side.
This is an approximation, as the full coefficients depend on the amplitude incident on both sides, but it has the advantage of making our equations linear in the wave luminosity rather than the wave amplitude.
As a result we find
\begin{align}
\label{eq:Fjp}
	L_{j}^{+} &= T_{j-1,j} L_{j-1}^{+} + R_{j}^{-+} L_{j}^{-} + L_{s,j}^+\\
\label{eq:Fjm}
	L_{j}^{-} &= T_{j+1,j} L_{j+1}^{-} + R_{j}^{+-} L_{j}^{+} - L_{s,j}^+
\end{align}
where $T_{j,j+1}$ is the transmission coefficient across the region between boundaries $j$ and $j+1$, $R_j^{\pm \mp}$ is the reflection coefficient at boundary $j$ going from the $\pm$ direction to $\mp$, and $L_{s,j}^{\pm}$ is any new wave luminosity being sourced by a convection zone adjacent to boundary $j$, if there is one.
The details of how we compute $T$, $R$, $L_{s}$, and $L_{\rm damp}$ are given in Appendix~\ref{appen:propagation}.

Equations~\eqref{eq:steady},~\eqref{eq:Fjp}, and~\eqref{eq:Fjm} form a tridiagonal linear system of equations for the wave luminosity at each boundary $j$, frequency $f$, and quantum number $\ell$.
We solve this numerically\footnote{As a technical note, the linear system of equtions may be nearly singular for modes with very little damping. We therefore place an upper bound on the damping length in each propagating region of $10^8 \Delta r$, where $\Delta r$ is the region width. This ensures that a well-conditioned solution can be found.} and then calculate the wave luminosity everywhere inside each propagating region using WKB.
We do not calculate the wave luminosity explicitly inside evanescent regions because we never directly use it there.

Note that this procedure requires us to assume that the waves are in steady state, so that the energy dissipated as heat equals the energy input from equation~\eqref{eq:flux_avg}.
This assumption is good for most modes most of the time, but fails during the $\sim 1\mathrm{yr}$ peak of the flash, which is shorter than the damping times of some high-frequency low-$\ell$ modes.
For those modes the dissipation should be spread out over a time-scale longer than the flash.
We do not think this changes our results qualitatively, because most of the power is in modes which dissipate rapidly during the flash, and because the phenomenology in our models is not sensitive to the precise timing of the wave heating, but if the peak frequency of the spectrum were substantially higher than we have assumed this could become a problem.

\subsection{Wave Heating}

Radiative diffusion damps the IGW amplitude on the length-scale~\citep{Fuller2014}
\begin{align}
	L_{\rm d} = \frac{2 \omega N^2}{N_{\rm T}^2 \alpha k_r k^2},
	\label{eq:Ldg}
\end{align}
where
\begin{align}
	\alpha = \frac{16 \sigma T^3}{3\kappa \rho^2 c_p}
\end{align}
is the radiative thermal diffusivity and $\boldsymbol{k}$ is the wave-vector.
When IGW tunnel through an evanescent region to become p-modes, they instead damp on the length-scale
\begin{align}
	L_{\rm d} = \frac{2 \omega}{\alpha k_r k^2 \nabla_{\rm ad}},
	\label{eq:Ldp}
\end{align}
where $\nabla_{\rm ad}$ is the adiabatic temperature gradient.

See Appendix~\ref{appen:damping} for a derivation of equation~\eqref{eq:Ldp} and a discussion of how to smoothly interpolate between equations~\eqref{eq:Ldg} and~\eqref{eq:Ldp}.

We neglect radiative damping in evanescent zones because the waves which carry most of the energy typically have very long wave-lengths in evanescent zones.
Moreover in certain evanescent zones (i.e. convection zones) the entropy profile is nearly adiabatic, so the temperature perturbation of the waves is small and the damping rate is reduced relative to propagating regions.

Damping produces the specific wave heating
\begin{align}
	\epsilon_{\rm wave} &= \frac{2 L_{\rm wave}}{4\pi r^2 \rho L_{\rm d}},
	\label{eq:eps_wave}
\end{align}
where $L_{\rm wave}$ is the unsigned sum of the wave luminosity travelling both up and down at any given point in the star\footnote{That is, $L_{\rm wave} = L_{\rm wave,up} + L_{\rm wave,down}$ and is always a positive number.}.

\subsection{Numerical Implementation}

In each MESA timestep we identify all convection zones.
We then calculate the waves these source on a grid of $\omega$ and $\ell$ by integrating the spectrum (Equation~\ref{eq:LQ13}) through each zone and attenuating them according to the evanescent wavevectors.
Waves are then propagated through the model as described in Section~\ref{sec:prop}, and the heating is calculated via equation~\eqref{eq:eps_wave}.
This heat appears as an additional source term in the equation of energy conservation which MESA solves.

\section{Stellar Models} \label{sec:models}

\subsection{Canonical Model}

\begin{figure*}
\centering
\includegraphics[width=0.95\textwidth]{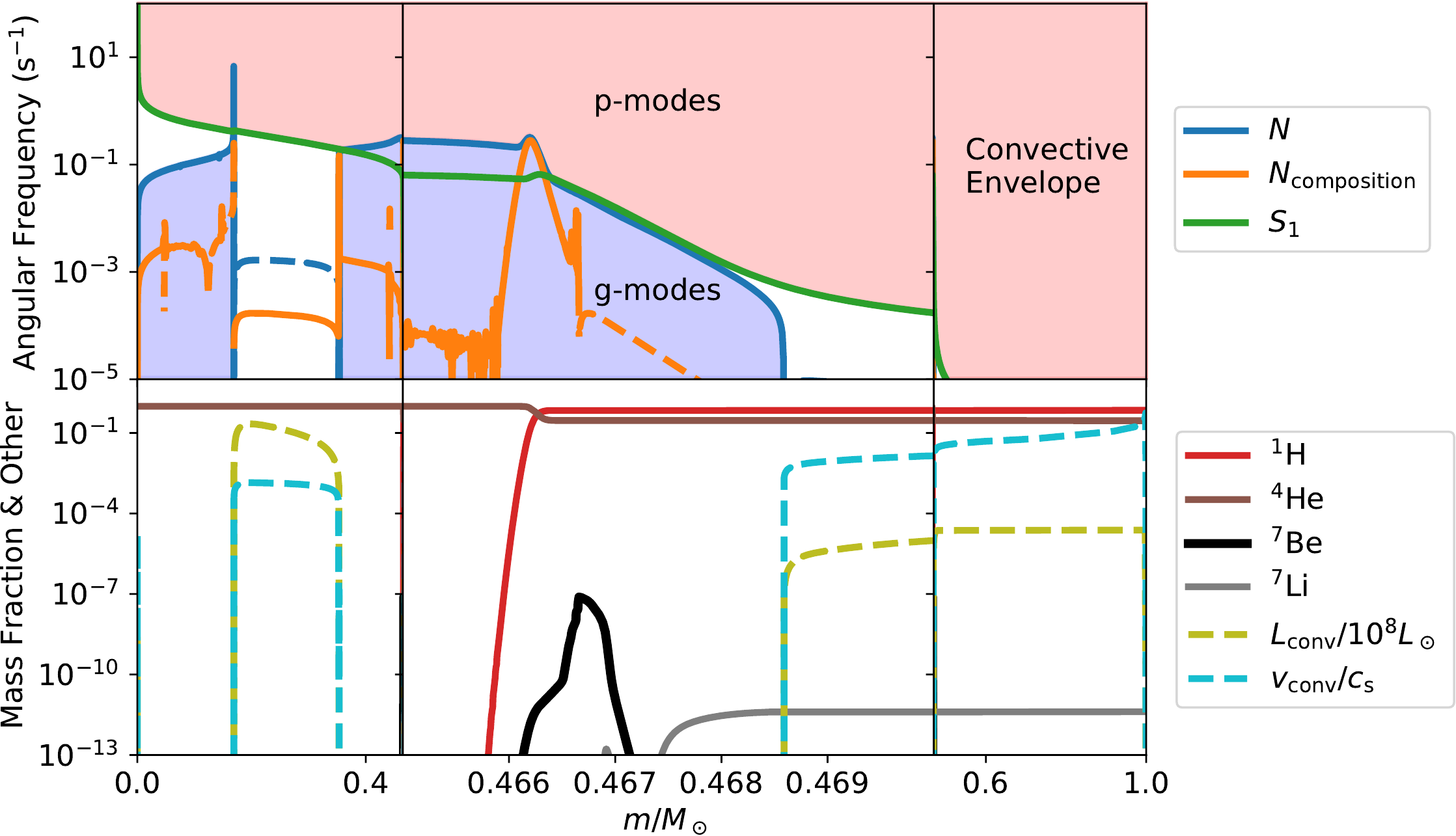}
\caption{A model taken from shortly before the peak of the helium flash, with nuclear luminosity $L_{\rm nuc} \approx 3\times 10^7 L_\odot$. Quantities are shown as functions of mass coordinate. Note that the three panels have different horizontal scales so that the middle one emphasizes the hydrogen burning shell. (Upper) A propagation diagram showing the \brvs\ frequency $N$, the contribution due to composition $N_{\rm composition}$, and the Lamb frequency for $l=1$ modes ($S_1$) as functions of mass coordinate $m$. (Lower) The convective luminosity $L_{\mathrm{conv}}$, convective Mach number $v_{\rm c} / c_{\rm s}$, and abundances of different chemical species are shown as a function of mass coordinate.}
\label{fig:schema2}
\end{figure*}

We begin with a model on the zero-age main-sequence and evolve it through the flash, calculating the wave heating and adding that to the MESA energy equation.
We use wave power multiplier $\beta=5$ to show an optimistic case with lots of wave heating.
Figure~\ref{fig:schema2} shows the structure of this model early into the flash, at a point where $L_{\rm nuc} \approx 3\times 10^7 L_\odot$.
At this stage, before wave heading has had a substantial impact, all of our models have a similar structure.

The upper panel shows a propagation diagram, while the lower panel shows the abundances of several key species as well as the convective Mach number and luminosity.
In the core, helium burning has produced a convection zone (the HeCZ) spanning approximately $0.2 M_\odot$.
At the peak of the flash the HeCZ extends further up, its Mach number reaches nearly $0.1$, and its luminosity peaks above $3\times 10^9 L_\odot$.
This convection zone therefore temporarily emits of order $L_{\rm wave} \approx \mathcal{M} L_{\rm conv} \approx 3\times 10^8 L_\odot$ of IGW\footnote{The wave luminosity we would calculate with $\beta=1$ is somewhat lower than this because the HeCZ is not homogeneous, and the typical convective luminosity and Mach number are somewhat smaller than their peak values. In our calculations we boost the wave luminosity with $\beta=5$, and this makes $L_{\rm wave}$ peak around $8\times 10^7 L_\odot$.}.

\begin{figure}
\centering
\includegraphics[width=0.47\textwidth]{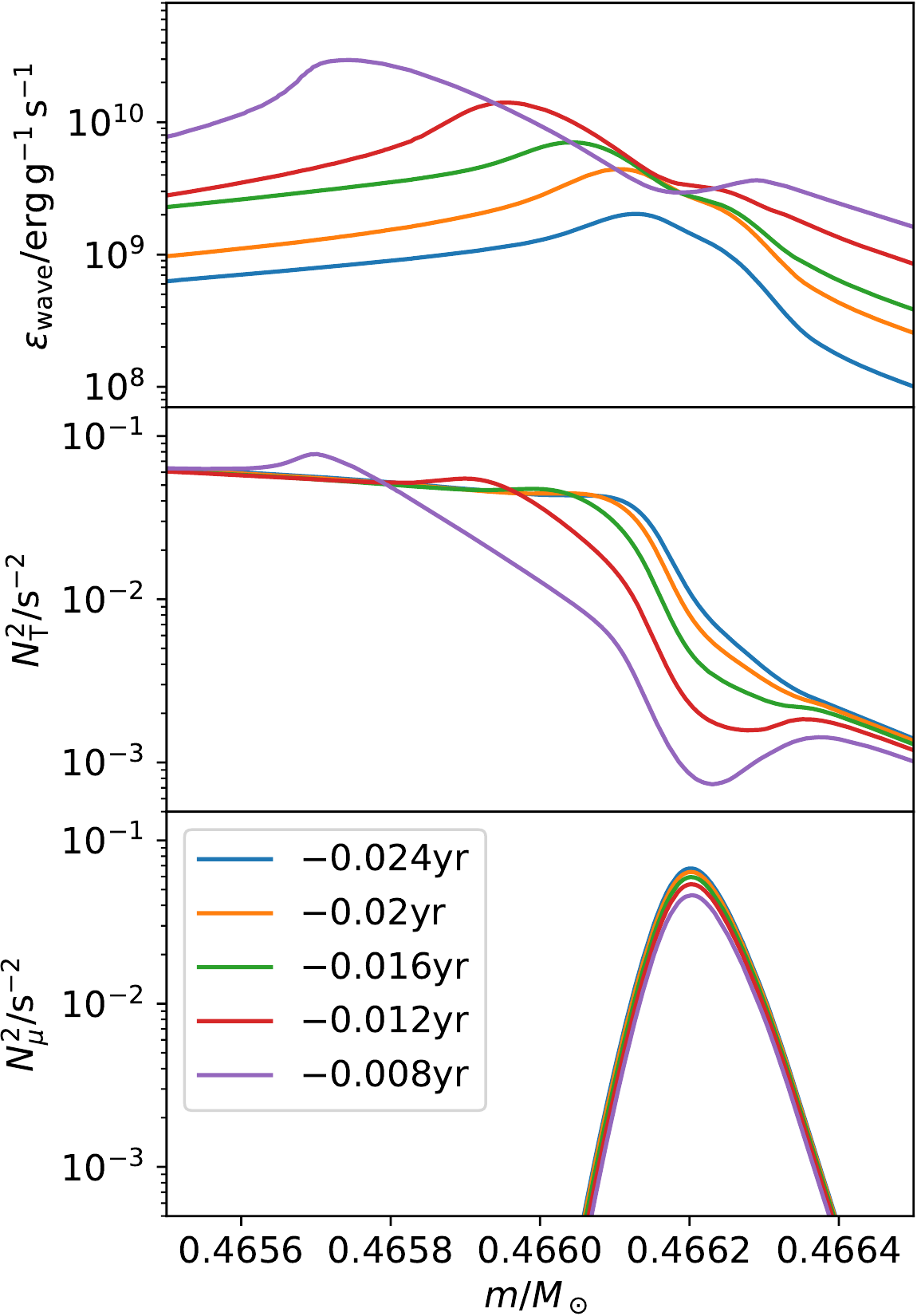}
\caption{Profiles of $\epsilon_{\rm wave}$, the thermal \brvs\ frequency $N_{\rm T}^2$, and the compositional \brvs\ frequency $N_\mu^2$ are shown for models shortly before the peak of the helium flash.}
\label{fig:brunt}
\end{figure}

Figure~\ref{fig:brunt} shows profiles of the resulting wave heating, thermal \brvs\ frequency, and compositional \brvs\ frequency.
These are taken from snapshots shortly before the peak of the flash.
Waves travelling upwards from the HeCZ reach the hydrogen burning shell at a mass coordinate of $m \approx 0.466 M_\odot$, where the composition gradient causes a maximum in the \brvs\ frequency $N^2$.
This shortens the mode wavelengths and increases damping, so most of the wave power is deposited near the hydrogen burning shell as heat.

Because the time-scale of the flash is short compared with the local thermal time the heat does not have time to diffuse away, so the entropy rises at the point of maximum wave heating.
This gradually makes the entropy gradient steeper on the inner edge of the maximum and shallower on the outer edge, so $N^2$ increases below the maximum and the heating gradually shifts deeper into the model.
Starting around $10^{-1}\rm yr$ before the peak of the flash, enough heat has accumulated to produce a second maximum in $N^2$ which detaches from the composition maximum, and this new thermal $N^2$ feature gradually moves downwards through the model.

\begin{figure*}
\centering
\includegraphics[width=0.96\textwidth]{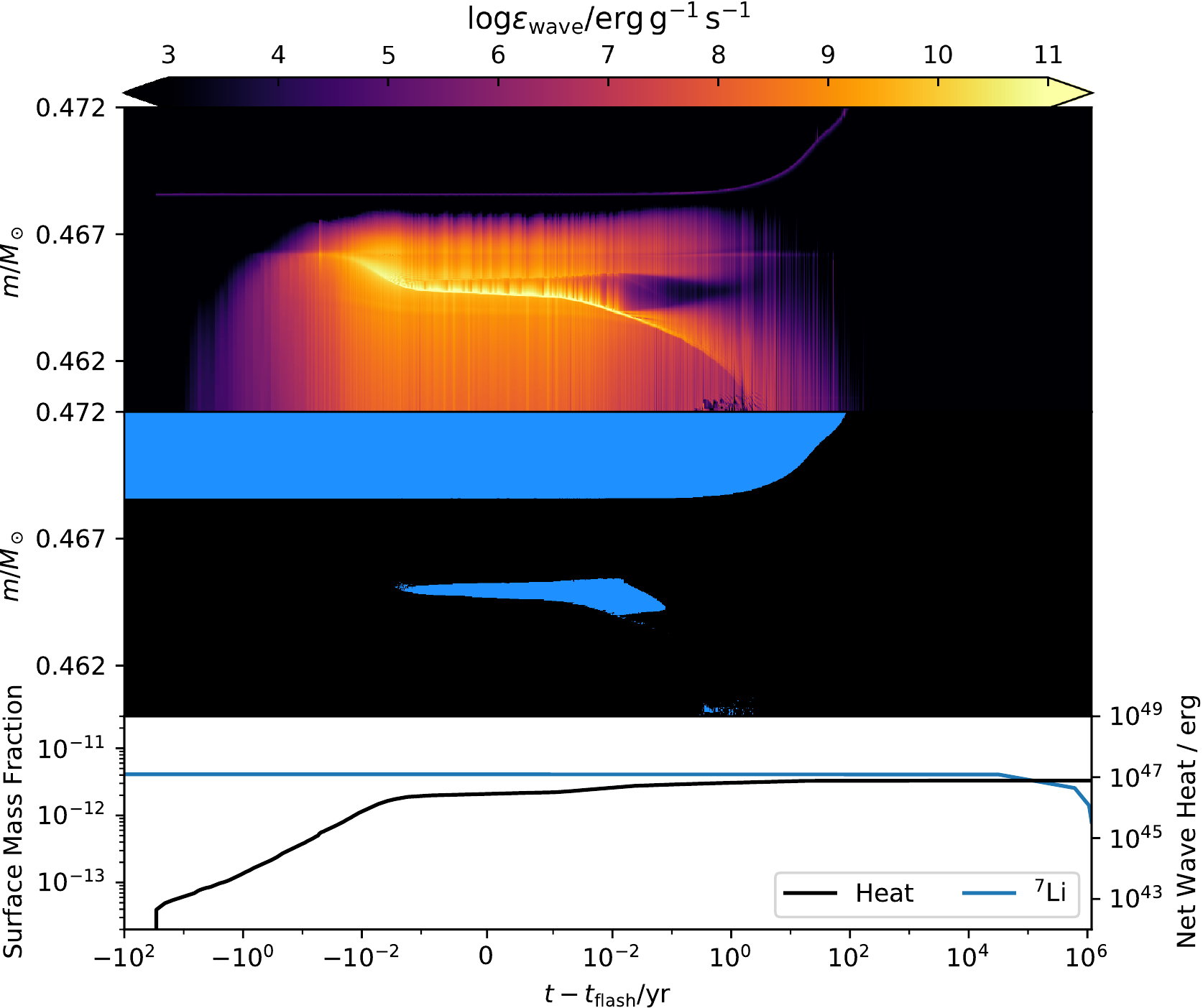}
\caption{The time evolution of a Red Giant through the helium flash is shown for a model with heat multiplier $\beta = 5$. (Upper) The specific wave heating rate $\epsilon_{\rm wave}$ is shown as a function of mass (vertical) and time (horizontal). Time is shown on a logarithmic scale referenced to the peak of the flash, becoming linear for the $10^{-3}\mathrm{yr}$ period on either side of the flash. (Middle) A Kippenhahn diagram shows convective regions as blue. (Lower) Surface abundances of $^{7}{\rm Be}$ and $^{7}{\rm Li}$, as well as the net wave heat deposited above $m/M_\odot=0.466$ are shown as functions of time.}
\label{fig:kipp}
\end{figure*}

This may also be seen in the Kippenhahn diagram in Figure~\ref{fig:kipp}.
Over time the heating shifts deeper and intensifies.
Starting around $10^{-2}\rm yr$ before the peak of the flash, the entropy at the point of maximum heating has risen so much that the gradient above the peak changes sign and becomes convectively unstable.
This results in the formation of a convection zone just beneath the hydrogen burning shell.
Because the convection zone is evanescent to IGW, it serves as a barrier of sorts, reducing the amount of heat which makes it through to the shell.

\subsection{What goes wrong?}

Our hope, outlined in Section~\ref{sec:story}, was that the heating would remain peaked around the hydrogen burning shell, form a convection zone there, and eventually extend that new convection zone upwards to meet the convective envelope.
The reason we do not see this in our models is that the heating always shifts downwards from the hydrogen burning shell, which buries the newly-formed convection zone too deep to ever reach the convective envelope.

We have tested a number of possible changes to the input physics to see if there is a combination which produces the behavior we originally expected.
These included varying the wave luminosity multiplier $\beta$, varying rotation rates from non-rotating (as in our canonical model) to near-breakup, varying the slope of the wave spectrum in both $\omega$ and $\ell$, and a wide variety of convergence tests.
No combination we have tried so far produces the behavior described in Section~\ref{sec:story}.

\section{Discussion and Conclusions} \label{sec:conclusions}

We have examined whether or not heating by internal gravity waves can convectively mix beryllium from the hydrogen-burning shell into the envelope during the helium flash.
We have found that this mechanism does not operate for reasonable wave luminosities because of a subtle interaction between the entropy gradient and the wave heating, which results in the heating shifting down from the hydrogen burning shell during the course of the flash.

Nonetheless, there are significant uncertainties in the details of wave heating in these stars, and it is conceivable that physics we have not considered could make this mechanism operate.
If our mechanism operates, what predictions does that make?

First, stars cannot be enhanced by this mechanism beyond $A(\mathrm{Li}) \approx 1.5$, because doing so requires more beryllium than is present in the hydrogen-burning shell.
This means that it cannot explain the small population of very Li-rich giants with $A(\mathrm{Li}) > 2$.
Those objects likely need to be formed via other channels such as binary interactions~\citep{1999MNRAS.308.1133S,2013MNRAS.430.2113Z,2019ApJ...880..125C} or planetary engulfement~\citep{2020arXiv200205275S}.

\citet{Schwab_2020} was able to produce dramatic enhancement of lithium, up to $A(\mathrm{Li}) \approx 4$, by using very long mixing events ($\sim 10^4 \mathrm{yr}$).
In our models the IGW luminosity is only significant for $\sim 10\mathrm{yr}$, set by the nuclear evolution of the helium flash.
The difference between these scenarios is that with very brief mixing only the beryllium that already exists in the hydrogen burning shell is available for dredge-up, whereas with more extended mixing the burning shell can enhance the entire mixed region up to the pp-II equilibrium $^{7}\rm Be$ abundance.
As a result the two scenarios differ in their $^{7}\rm Li$ enhancement by a factor of the ratio of the envelope mass to the burning shell mass, which is of order $300$.
That is, if our mechanism were to operate it could only produce enhancement up to $A(\mathrm{Li}) \approx 1.5$, and other physics such as sustained rotational mixing following tidal spin-up and contraction onto the clump~\citep{2019ApJ...880..125C} would be needed to account for super Li-rich clump stars.

Additionally, our mechanism can only operate in stars which experience the helium flash, and so cannot operate in more massive stars ($M \ga 2 M_\odot$) and generally produces less $^{7}\rm Li$ enhancement towards higher masses as the flash weakens.
Hence, if our mechanism operates it could account for the low-mass ($M \lesssim 2 \, M_\odot$) moderately Li-rich clump stars of~\citet{2020MNRAS.494.1348D,2021MNRAS.tmp.1175D,2021ApJ...913L...4S,2021NatAs...5...86Y}, but not for the high-mass Li-rich clump stars in their samples.
The fact that high-mass clump Li-rich stars do exist suggests that different mechanisms (e.g., binary interactions) are responsible for the very Li-rich phenomenon.

Finally, in our scenario lithium remains enhanced for the duration of the Red Clump.
This evolution is not intrinsic to our model, however, and just reflects the lack of any other (slower) mixing mechanism which can carry lithium from the envelope to hotter regions where it burns.
Other processes which we have neglected, such as rotational mixing, could play an important role in destroying lithium during the longer-term post-flash evolution of Red Clump stars.
Lacking such processes, at the onset of the AGB we see the lithium abundance falls by $\sim 0.5\rm dex$ due to thermohaline mixing pulling lithium down into hot burning regions.

\acknowledgments

The Flatiron Institute is supported by the Simons Foundation.
A.S.J. thanks the Gordon and Betty Moore Foundation (Grant GBMF7392) and the National Science Foundation (Grant No. NSF PHY-1748958) for supporting this work.
J.F. is thankful for support through an Innovator Grant from The Rose Hills Foundation, and to the Sloan Foundation through grant FG-2018-10515.
We are grateful to Yuri Levin and Josiah Schwab for helpful conversations on this topic, and to Lucy Reading-Ikkanda from the Simons Foundation for producing Figure~1.

\software{
\texttt{MESA} \citep[][\url{http://mesa.sourceforge.net}]{Paxton2011,Paxton2013,Paxton2015,Paxton2018,Paxton2019},
\texttt{MESASDK} x86\_64-linux-20.8.2 \citep{mesasdk_linux,mesasdk_macos},
\texttt{matplotlib} \citep{hunter_2007_aa}, 
\texttt{NumPy} \citep{der_walt_2011_aa}, and
         }

\onecolumngrid 

\clearpage
\appendix

\section{Wave Propagation Details} \label{appen:propagation}

Here we calculate the coefficients of transmission ($T$) and reflection ($R$), the wave luminosity lost to damping ($L_{\rm damp}$), and the wave excitation $L_{s}$.

\subsection{Propagating Regions}\label{sec:prop}

Consider a propagating region between boundaries $j$ and $j+1$ with luminosity $L_{j}^{+}$ entering through boundary $j$ heading downward towards boundary $j+1$.
This luminosity propagates according to
\begin{align}
	\frac{dL}{dr} = \frac{2 L}{L_{\rm d}},
	\label{eq:propp}
\end{align}
decaying away from boundary $j$, eventually giving luminosity at boundary $j+1$ leaving the region
\begin{align}
	F = L_{j}^{+} e^{-\int \frac{2 dr}{L_{\rm d}}}.
\end{align}	
The transmission factor across the propagating region is then
\begin{align}
	T_{j,j+1} = e^{-\int \frac{2 dr}{L_{\rm d}}}.
\end{align}

Note that there is a symmetry here: had we examined luminosity entering through boundary $j+1$ travelling upwards we would pick up a minus sign in the differential equation but also swap the direction of integration, so the transmission factor would be the same. That is,
\begin{align}
	T_{j,j+1} = T_{j+1,j}
\end{align}

The energy which is not transmitted across the propagating region is lost to damping, so
\begin{align}
	L_{\rm damp}^{j,j+1} = (1 - T_{j,j+1}) (L_{j}^{+} + L_{j+1}^{-}).
	\label{eq:Fdamp}
\end{align}

Finally, the reflection coefficient for waves \emph{entering} a propagating region we set to $R_{j}^{+-} = R_{j+1}^{-+} = 1$.
This is a mathematical convention which requires that the transmission coefficient we compute for evanescent regions incorporate the effects of reflection at both boundaries of the evanescent region.
This turns out to be mathematically simpler than treating evanescent and propagating regions on equal footing because the wave luminosity is not well-defined in evanescent regions.

\subsection{Evanescent Regions}

\begin{figure}
\centering
\includegraphics[width=0.7\textwidth]{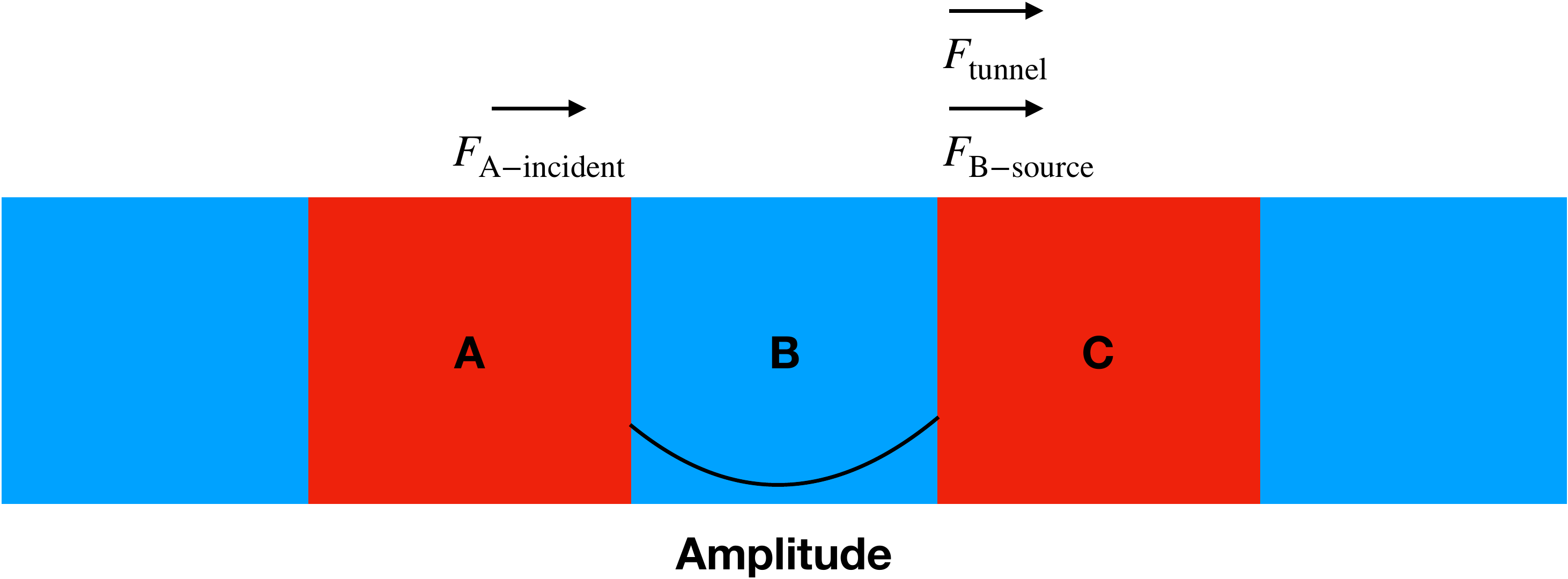}
\caption{A sequence of evanescent (blue) and propagating (red) regions are shown. $L_{\mathrm{A}}^{-}$ is the total luminosity incident on turning point $\mathrm{AB}$. Some of this tunnels through to turning point $\mathrm{BC}$ and emerges as $L_{\rm tunnel}$. In addition, convection in region $\mathrm{B}$ sources wave luminosity $L_{B-\rm source}$ emerging at turning point $2$.}
\label{fig:schema0}
\end{figure}

Processing an evanescent region is more complicated.
Some of the key quantities involved are shown in Figure~\ref{fig:schema0}.
We separate this process into several tasks:
\begin{enumerate}
	\item Compute the fraction of the luminosity from the \emph{previous} propagating region which tunnels through to the \emph{next} propagating region.
	\item Accumulate any sources of IGW excitation through the evanescent region, accounting for the decay of the IGW wavefunctions through the region.
	\item Compute the fraction of that excited luminosity which escapes to the next propagating region.
\end{enumerate}
These complications do not arise in propagating regions because our only sources are convection zones, which are always evanescent for IGW, and because waves travel rather than tunnelling through propagating regions.

\subsubsection{Tunnelling} \label{sec:tunnelling}

We handle (1) by computing the transmission factor across the evanescent region within the WKB approximation, where the wave eigenfunctions take the form
\begin{align}
	\xi_r = \sqrt{\frac{k_\perp^2}{r^2 \rho k_r}}\left(\alpha e^{i \int k_r dr - i \omega t} + \beta e^{i \int k_r dr + i \omega t}\right).
	\label{eq:eigen}
\end{align}
The factor of $\sqrt{k_\perp^2/r^2 \rho k_r}$ ensures that the wave luminosity is conserved, because
\begin{align}
	F &= \frac{1}{2} \rho \frac{\omega}{k_r} u^2= \frac{1}{2} \rho \frac{\omega^3}{k_r} \xi^2\approx \frac{1}{2} \rho \frac{\omega^3}{k_r} \xi_\perp^2\approx \frac{1}{2} \rho \frac{\omega^3 k_r}{k_\perp^2} \xi_r^2,
\end{align}
where we have used the fact that for g-modes $\xi_r k_r \approx u_\perp k_\perp$ and $k_r \gg k_\perp$, which means that $\xi \approx \xi_\perp \approx \xi_r k_r / k_\perp$.
Using the fact that $r k_\perp$ is a constant in $r$, we re-express equation~\eqref{eq:eigen} as
\begin{align}
	\xi_r = \frac{1}{\sqrt{r^4 \rho k_r}}\left(\alpha e^{i \int k_r dr - i \omega t} + \beta e^{i \int k_r dr + i \omega t}\right),
	\label{eq:eigen2}
\end{align}
which is equivalent up to rescaling by a factor which depends only on $\ell$.

\begin{figure}
\centering
\includegraphics[width=0.7\textwidth]{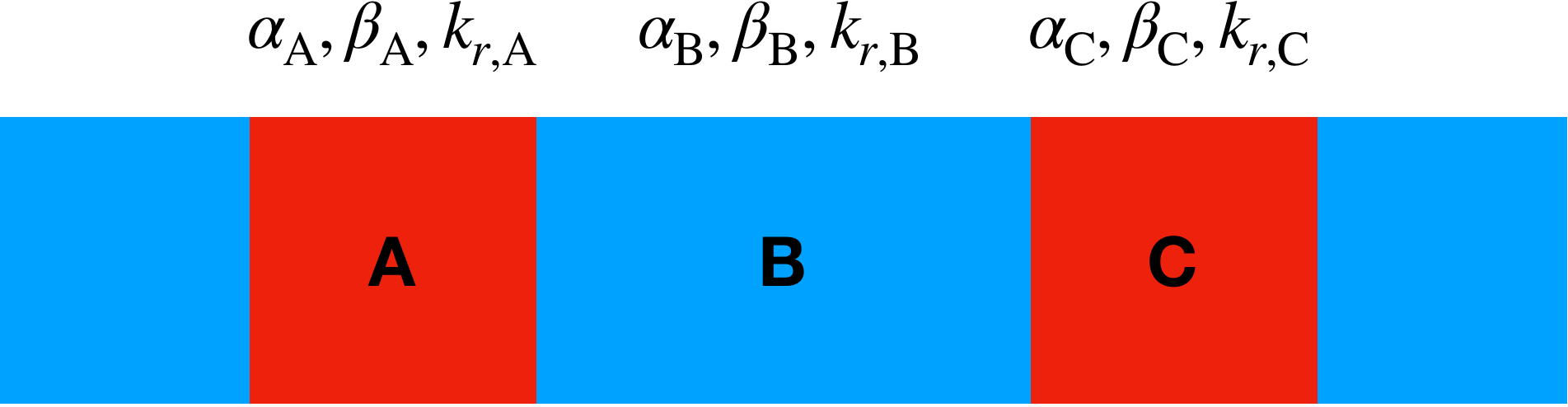}
\caption{A sequence of evanescent (blue) and propagating (red) regions are shown along with the relevant WKB quantities for computing tunnelling coefficients.}
\label{fig:schema1}
\end{figure}

Call the previous propagating region A, the current evanescent region B, and the next propagating region C, as shown in Figure~\ref{fig:schema1}.
Within each region the wave amplitude can be written with equation~\eqref{eq:eigen2}.
We label all quantities in this equation by their respective regions.
We fix $\alpha_{\rm A} = 1$ and $\beta_{\rm C}=0$ to represent a wave incident from region A onto region B.
Because the dispersion relation (equation~\ref{eq:dispersion}) is quadratic in $k_r$ the eigenfunction $\xi_r$ and its first derivative must be continuous at the turning points.
For the waves of interest $h k_r \gg 1$, so we neglect the derivatives of $k_r$ and $\rho$ and write these conditions as
\begin{align}
	1 + \beta_{\rm A} &= \left(\alpha_{\rm B} + \beta_{\rm B}\right)\sqrt{\frac{k_{r, \rm A}}{k_{r, \rm B,L}}},\\
	k_{r,\rm A}\left(1 - \beta_{\rm A}\right) &= k_{r,\rm B,L}\left(\alpha_{\rm B} - \beta_{\rm B}\right)\sqrt{\frac{k_{r, \rm A}}{k_{r, \rm B,L}}},\\
	\alpha_{\rm B} e^{-\int_{r_{\rm A-B}}^{r_{\rm B-C}} k_{r,\rm B} dr} + \beta_{\rm B} e^{\int_{r_{\rm A-B}}^{r_{\rm B-C}} k_{r,\rm B} dr} &= \alpha_{\rm C}\sqrt{\frac{k_{r, \rm B,R}}{k_{r, \rm C}}},\\
	k_{r,\rm B,R}\left(\alpha_{\rm B}e^{-\int_{r_{\rm A-B}}^{r_{\rm B-C}} k_{r,\rm B} dr} - \beta_{\rm B}e^{\int_{r_{\rm A-B}}^{r_{\rm B-C}} k_{r,\rm B} dr} \right) &= \alpha_{\rm C}k_{r,\rm C}\sqrt{\frac{k_{r, \rm B,R}}{k_{r, \rm C}}},
\end{align}
where the labels A-B and B-C are quanties evaluated at the interface between regions A-B and B-C respectively, and the labels L/R denote quantities evaluated on the left or right-hand side of region B.
Solving for $\alpha_{\rm C}$ and noting that the wave luminosity in the forward direction ($\mathrm{A} \rightarrow \mathrm{C}$) scales as $\alpha^2$ we obtain the transmission coefficient
\begin{align}
	T_{j,j+1} = \left|\frac{\alpha_{\rm C}}{\alpha_{\rm A}}\right|^2 = \left|\alpha_{\rm C}\right|^2 = \frac{4 k_{r,\rm A} |k_{r,\rm B,L} k_{r,\rm B,R}| k_{r,\rm C}}{\sinh^2(\lambda)\left(|k_{r,\rm B,L} k_{r,\rm B,R}|^2 - k_{r,\rm A} k_{r,\rm C}\right)^2 + \cosh^2(\lambda) \left|k_{r,\rm B,R}k_{r,\rm A}+k_{r,\rm B,L}k_{r,\rm C}\right|^2},
	\label{eq:transmission}
\end{align}
where
\begin{align}
	\lambda \equiv \int_{r_{\rm B}}^{r_{\rm C}} \mathrm{Im}(k_r) dr.
\end{align}
In calculating $\mathrm{Im}(k_r)$ we account for both radiative damping and evanescent decay, combining these as $\mathrm{Im}(k_r) = \mathrm{Im}(k_{r,\rm decay}) + L_{\rm d}^{-1}$.
Note that $\cosh(\lambda)$ and $\sinh(\lambda)$ are both proportional to $e^{\lambda}$ for $\lambda \gg 1$, so in the limit of a thick evanescent zone the transmission factor decays as $T \propto e^{-2\lambda} = e^{-\int 2 \mathrm{Im}(k_r) dr}$.
Further note that $T_{j,j+1} = T_{j+1,j}$, as can be seen by swapping the A/C labels in equation~\eqref{eq:transmission}.

Note that equation~\eqref{eq:transmission} is only valid for tunnelling from g-modes to g-modes.
For other tunnelling avenues (e.g. g-modes to p-modes) a different expression is needed because the flux for p-modes has a different relation to the displacement.
We have assumed in our calculations that all tunnelling is from g-modes to g-modes, but it would be straightforward to extend our formalism to account for tunnelling from g-modes to p-modes too.

Because $k_r$ vanishes at the turning points we cannot evaluate $k_{r,\rm A/B/C}$ in equation~\eqref{eq:transmission} precisely at these points.
Instead, we evaluate $k_{r,\rm A/C}$ by walking away from each turning point into the adjacent propagating region until $|k_r^{-3} d k_r^2/dr| < 1$, which is an approximate test for when the wave-vector is sufficiently uniform that WKB is valid once more.
We cannot always obtain a condition like this for $k_{r,\rm B}$, so in region B we use $k_{r,\rm B} = \lambda / \Delta r$, where $\Delta r$ is the width of the evanescent region.

\subsubsection{Excitation} \label{sec:excitation}

We compute the contribution of each cell in the model to the wave excitation.
To do so we follow the reasoning of~\citet{2013MNRAS.430.2363L} with two modifications.
First, because we explicitly calculate the transmission of waves from the convection zone into the radiative zone within WKB we need to omit the transmission factor they assume.
Secondly, because we perform the integration of the source against the mode eigenfunctions we need to calculate a source per unit distance in the convection zone, rather than an overall source as they do.

For the case of discontinuous $N^2$,~\citet{2013MNRAS.430.2363L} obtain
\begin{align}
\frac{dF}{d\log \omega d\log k_\perp} = \rho_0 v_{\rm c}^3 \mathcal{M} (k_\perp h)^4 \left(\frac{\omega}{\omega_c}\right)^{-13/2},
\end{align}
where $\mathcal{M}$ is the convective Mach number.
They derive this by assuming that eddies only contribute if they are within a distance $1/k_\perp$ of the boundary. We want to get rid of that factor because we explicitly integrate the source, weighting by the distance to the boundary using the WKB eigenfunctions, so we instead write
\begin{align}
\frac{dF}{dr d\log \omega d\log k_\perp} = \rho_0 v_{\rm c}^3  h^{-1} \mathcal{M} (k_\perp h)^5 \left(\frac{\omega}{\omega_c}\right)^{-13/2} - 2 \mathrm{Im}(k_r)\frac{dF}{d\log \omega d\log k_\perp},
\end{align}
where $L_{\rm d}$ is the evanescent decay length of the eigenfunctions.
Next we want to divide by their assumed transmission factor at the convective-radiative boundary, which is $\omega/N_0$, to obtain a source.
Thus we find
\begin{align}
\frac{dS}{dr d\log \omega d\log k_\perp} = \rho_0 v_{\rm c}^3 h^{-1} (k_\perp h)^5 \left(\frac{\omega}{\omega_c}\right)^{-15/2} - 2 \mathrm{Im}(k_r)\frac{dS}{d\log \omega d\log k_\perp}
\end{align}
We take $S=0$ at the boundary with the previous propagating region and then integrate this equation to find the source entering the next propagating region.

So far we have determined the scaling of the spectrum, but not its amplitude.
We normalize the source so that
\begin{align}
	\int S(r,f,\ell) d\log f d\log k_\perp = \beta L_{\rm conv},
	\label{eq:norm}
\end{align}
where $\beta$ is a free parameter of order unity we introduce because both estimates of the power are done to order-of-magnitude.
This normalization was chosen so that, after propagating the source into the radiative zone with a transmission factor of $\omega/N_{\rm propagating} \sim \mathcal{M}$, we match the expected emerging luminosity calculated by~\citet{1990ApJ...363..694G,2013MNRAS.430.2363L}.

Using the normalization from equation~\eqref{eq:norm} and recalling that the emerging luminosity is related to the source by a factor of order $\omega/c_s k_\perp$ in the Boussinesq limit (see Appendix~\ref{sec:escape}), we find that the total wave luminosity emitted in each direction by a deep convection zone is
\begin{align}
	L \approx \beta L_{\rm conv} \mathcal{M}_{\rm conv},
	\label{eq:luminosity_avg}
\end{align}
where $\mathcal{M}_{\rm conv}$ is now averaged over the convection zone using the evanescent wave eigenfunctions.

Equation~\eqref{eq:luminosity_avg} is not exact because the transmission factor is somewhat more complicated than the scaling $\omega/k_\perp$ suggests, depending on the full dispersion relation evaluated on either side of the boundary, and we account for those details in full in computing the emerging luminosity.
Moreover we use the local sound speed in equation~\eqref{eq:norm} and that may differ somewhat from the sound speed at the convective boundary, which introduces further approximation.
Despite this, in practice we find that the integrated spectrum is typically within factor of $2$ of $\beta L_{\rm conv} \mathcal{M}_{\rm conv}$, though the direction and magnitude of the difference varies with details of the structure of the star and can be larger in specific circumstances.

\subsubsection{Escape} \label{sec:escape}

Some of the source we accumulate in the evanescent region escapes to the next propagating region.
Following Section~\ref{sec:tunnelling} we compute this transmission factor by enforcing continuity of the eigenfunction and its first radial derivative at the boundary turning point.
Call the evanescent region B and the propagating one C.
Pick $r_{\rm B}$ and $r_{\rm C}$ to equal the boundary radius.
Assuming a unit source incident on the boundary we find
\begin{align}
	1 + \beta_{\rm B} &= \alpha_{\rm C},\\
	k_{r,\rm B}\left(1 - \beta_{\rm B}\right) &= k_{r,\rm C}\alpha_{\rm C}.
\end{align}
This produces the transmission factor
\begin{align}
 T = \left|\frac{\alpha_{\rm C}}{\alpha_{\rm B}}\right|^2 = \left|\alpha_{\rm C}\right|^2 = \frac{4 |k_{r,\rm B} k_{r,\rm C}|}{|k_{r,\rm B}+k_{r,\rm C}|^2}.
\end{align}
Note that because region B is evanescent we know that $k_{r,\rm B}$ is imaginary, and because region C is propagating we know that $k_{r,\rm C}$ is real. Hence
\begin{align}
 T = \frac{4 |k_{r,\rm B} k_{r,\rm C}|}{|k_{r,\rm B}|^2+|k_{r,\rm C}|^2}.
\end{align}
As before we evaluate both $k_{r,\rm B}$ and $k_{r, \rm C}$ away from the boundary, at the first point where $|k_r^{-3} d k_r^2/dr| < 1$.

Note that if the evanescent region is a convection zone and we are studying a g-wave then $\omega \approx |N| \ll c_s k_\perp$.
If, further, $\omega_{\rm ac} \ll c_s k_\perp$ (the Boussinesq limit) then $k_{r, \rm B} \approx i k_\perp$.
If the propagating region is a radiative zone then the dispersion relation (equation~\eqref{eq:dispersion}) gives $k_{r, \rm C} \approx (N_{\rm propagating} / \omega)  k_\perp \gg |k_{r,\rm B}|$.
So 
\begin{align}
	T \approx 4 \frac{k_{r, \rm B}}{k_{r, \rm C}} \propto \frac{\omega}{N_{\rm propagating}},
\end{align}
which is the scaling we used in Section~\ref{sec:excitation}.

\section{Radiative Diffusion} \label{appen:damping}

The wave velocity amplitude $u$ decays on the damping length-scale
\begin{align}
L_d = -\frac{dr}{d\ln u}
\end{align}
where here $u$ represents a root-mean-square averaged quantity over a wave cycle.
The specific wave energy is related to the amplitude by
\begin{align}
E = \frac{1}{2} u^2
\end{align}
so
\begin{align}
\frac{dE}{dr} = u \frac{du}{dr} = -\frac{u^2}{L_d} = -\frac{2 E}{L_d}
\end{align}
hence
\begin{align}
L_d = -2\frac{dr}{d\ln E}
\end{align}
If waves travel at a rate $v = \omega / k_r$ then
\begin{align}
L_d = -\frac{2\omega}{k_r} \frac{dt}{d\ln E}
\end{align}

Radiative damping only acts on the temperature field.  In the limit of slow damping the ratio of the peak energy in each different components (mechanical, thermal, gravitational, etc.) is fixed. Using this we find
\begin{align}
\frac{d\ln E}{dt} = \frac{1}{E} \frac{dE}{dt} = \frac{1}{E} \frac{dE_{\rm thermal}}{dt} = \frac{E_{\rm thermal}}{E}\frac{d\ln E_{\rm thermal}}{dt}
\end{align}
The temperature field diffuses at a rate
\begin{align}
\frac{d\delta \ln T}{dt} = -\alpha k^2
\end{align}
so
\begin{align}
\frac{d\ln E}{dt} = -\alpha k^2\frac{ E_{\rm thermal}}{E}
\end{align}
Hence
\begin{align}
L_d = \frac{2\omega E}{\alpha k^2 k_r E_{\rm thermal}}
\end{align}

\subsection{g-waves}

For g-waves a mechanical displacement generates a buoyant restoring force, which then generates motion. The buoyant restoring force is $N^2 \xi_r$, where $\xi_r$ is the radial displacement. This force has potential energy $N^2 \xi_r^2/2$. At the maximum displacement all of the energy in the wave lives in this potential. The thermal energy is the portion of this potential due to temperature perturbations, which is just $N_T^2 \xi_r^2/2$. So the ratio
\begin{align}
\frac{E_{\rm thermal}}{E} = \frac{N_T^2}{N^2}
\end{align}
hence we recover the usual expression~\citep{Fuller2014}
\begin{align}
L_d = \frac{2\omega N^2}{\alpha k^2 k_r N_T^2}
\end{align}

\subsection{p-waves}

For p-waves a mechanical displacement generates a pressure restoring force, which then generates motion. The pressure restoring force is
\begin{align}
\delta p = \xi k \rho\left.\frac{\partial p}{\partial \rho}\right|_s
\end{align}
This force has potential energy
\begin{align}
\left.\frac{1}{2} \xi^2 k^2 \rho \frac{\partial p}{\partial \rho}\right|_s
\end{align}
How much of this potential is due to the temperature perturbations rather than density perturbations? If the motion were isothermal (no temperature perturbations) the potential would instead be
\begin{align}
\frac{1}{2} \xi^2 k^2 \rho \left.\frac{\partial p}{\partial \rho}\right|_T
\end{align}
So the portion owing to temperature perturbations is
\begin{align}
\frac{E_{\mathrm{thermal}}}{E} = \frac{\partial p/\partial \rho|_s - \partial p/\partial \rho|_T}{\partial p/\partial \rho|_s} = \frac{\Gamma_1 - \chi_\rho}{\Gamma_1} = \nabla_{\rm ad},
\end{align}
where $\Gamma_1$ is the first adiabatic index, $\chi_\rho$ is the susceptibility at constant density, and $\nabla_{\rm ad}$ is the adiabatic temperature gradient.
So
\begin{align}
L_d = \frac{2\omega}{\alpha k^2 k_r \nabla_{\rm ad}}
\end{align}

\subsection{Interpolation}

In general there are really two restoring forces, one for pressure and one for buoyancy. If they work in-phase then the potential energy is
\begin{align}
\frac{1}{2}\xi^2 k^2 \rho\left.\frac{\partial p}{\partial \rho}\right|_s + \frac{1}{2}\xi_r^2 N^2
\end{align}
The ratio of thermal to total energy is then
\begin{align}
\frac{\xi_r^2 N_T^2 + \xi^2 k^2 \rho c_s^2 (\Gamma_1 - \chi_\rho)}{\xi_r^2 N^2 + \xi^2 k^2 \rho c_s^2 \Gamma_1}
\end{align}
This just smoothly transitions between the two limits, so we can more simply switch between the two prescriptions based the ratio $x \equiv \omega / k c_s$. i.e.
\begin{align}
L_d = \frac{2\omega}{\alpha k^2 k_r}\left(\frac{1}{\nabla_{\rm ad}} x + (1-x) \frac{N^2}{N_T^2}\right)
\end{align}
This is somewhat ad-hoc, but should capture the relevant scaling.

\section{MESA Implementation} \label{appen:mesa}

\subsection{Microphysics}

We calculate our stellar models using revision r15140 of the Modules for Experiments in Stellar Astrophysics
\citep[MESA][]{Paxton2011, Paxton2013, Paxton2015, Paxton2018, Paxton2019}.
Our starting models are the pre-flash models of~\citet{Schwab_2020} retrieved from~\citet{https://doi.org/10.5281/zenodo.3960434}.

The MESA EOS is a blend of the OPAL \citet{Rogers2002}, SCVH
\citet{Saumon1995}, PTEH \citet{Pols1995}, HELM
\citet{Timmes2000}, and PC \citet{Potekhin2010} EOSes.

Radiative opacities are primarily from OPAL \citep{Iglesias1993,
Iglesias1996}, with low-temperature data from \citet{Ferguson2005}
and the high-temperature, Compton-scattering dominated regime by
\citet{Buchler1976}.  Electron conduction opacities are from
\citet{Cassisi2007}.

Nuclear reaction rates are a combination of rates from
NACRE \citep{Angulo1999}, JINA REACLIB \citep{Cyburt2010}, plus
additional tabulated weak reaction rates \citet{Fuller1985, Oda1994,
Langanke2000}. 
Screening is included via the prescription of \citet{Chugunov2007}.
Thermal neutrino loss rates are from \citet{Itoh1996}.

\subsection{Stellar Models}

Our stellar models generally follow the implementation of~\citet{Schwab_2020}.

Mixing length theory was implemented following~\citet{1968pss..book.....C} with mixing length parameter $\alpha = 1.8$.
The Ledoux criterion was used to determine convective stability.
Thermohaline mixing was implemented following~\citet{1980A&A....91..175K} with efficiency parameter $\alpha_{\rm thermohaline} = 100$ following~\citet{Schwab_2020} and~\citet{2020NatAs.tmp..139K}.

We use the MESA \texttt{pp\_and\_cno\_extras} nuclear reaction network, which includes the pp-chains and CNO-cycles.
The default MESA rate for the crucial $^{7}\mathrm{Be}$ electron capture reaction does not include the effects of bound states, which matter greatly for the temperatures $T < 10^7\mathrm{K}$ prevalent in the convective envelope.
To remedy this we use rate calculations from~\citet{2013ApJ...764..118S} which incorporate bound states for this reaction.
These were made available in machine-readble form by~\citet{2019A&A...623A.126V}.

Our models are initialized on the pre-main sequence with $Z=0.014$ except where otherwise stated, and use the solar abundance pattern of~\citet{doi:10.1146/annurev.astro.46.060407.145222}.
These choices reproduce the meteoric $A(\mathrm{Li}) = 3.26$ on the pre-main sequence.

\section{Data Availability}

The final configuration files and code used in our model grids are available in~\citet{zenodo}.
These are given in \texttt{Python}~3 Pickle files which specify the changes to make to the configuration on top of a base configuration given in the file `inlist\_project'.
These Pickle files also provide the short-sha's of \texttt{git} commits which can be found in the git repository stored in~\citet{zenodo}.
Each such commit corresponds to a single \texttt{MESA} run directory used to perform one of our runs, including the full configuration files and `run\_star\_extras' code used.
We further provide the Pickle files specifying the configurations and short-sha's of commits we used in the final set of convergence tests which demonstrate that our results are converged.
Our canonical model has short-sha \texttt{16d4}.

The same \texttt{git} repository contains a history of nearly all \texttt{MESA} runs used to develop this work contributed to this work.
These are commits whose messages contain the word `patch' and which do not lie on any branch.
These \texttt{git} experiments were performed using the \texttt{RemoteExperiments} software package, details of which may be found at \url{https://github.com/adamjermyn/remote_experiments}.

\subsection{Wave Heating}\label{appen:mesa_wave}

At the beginning of each MESA time-step we compute the wave excitation, propagation and heating on a discrete frequency and $\ell$ grid.
The wave heating is added to the MESA energy equation as an extra energy source, which is taken to be fixed for the duration of the time-step.

Because our calculation is performed within WKB we neglect effects related to the derivative of $k_r$ with respect to $r$.
We believe that such effects likely manifest at least in part by smoothing the effective stellar structure seen by the waves.
 of $k_r$ seen by the waves.
That calculation is beyond the scope of this work, so we implemented a simple smoothing of $k_r$ which occurs on a length scale $1/q k_r$, where $q$ is an order-unity free parameter.
We first calculate $k_r$, then use this to smooth $\omega_{\rm ac}$, $N^2$, and the thermal component $N_T^2$ of the Br\"unt-V\"ais\"al\"a frequency.
We then use these to calculate a smoothed $k_r$.
We studied the effects of varying $q$ and found that the amount of smoothing does make a difference to the precise location the heat is deposited, which makes us wary of including too much smoothing.
Our canonical models use $q=3$, which represents a conservative amount of smoothing.

The frequency grid is geometrically spaced by factors of $e^{\Delta \ln \omega}$ from $10^{-2}|N|_{\rm min}$ to $10^2 |N|_{\rm max}$, where $|N|_{\rm min/max}$ are the minimum and maximum magnitudes of the \brvs\ frequency in the convection zones.
The $\ell$ grid is geometrically spaced by factors of $e^{\Delta \ln \ell}$ from $1$ to $10 \ell_{\rm max}$, where $\ell_{\rm max}$ is the maximum of $r/h$ in the model.
We performed a series of runs varying these resolution controls and holding them equal to one another.
We found convergence with respect to $\omega,\ell$ resolution for settings finer than $\approx 0.4$. For our canonical models we adopt a resolution of $\Delta \ln \omega = \Delta \ln \ell = 0.3$.

The MESA cells are distributed according to a custom mesh function that places more resolution near sharp gradients in the $^{1}{\rm H}$ and $^{7}{\rm Be}$ abundances.
The overall mesh resolution is controlled by the parameter `mesh\_delta\_coeff`.
We disable remeshing during the flash itself because the MESA remeshing procedure can produce artefacts in the $N^2$ profile which then produce numerically unstable artifacts in the wave heating profile.

We performed a series of runs attempting to converge our results with respect to mesh resolution.
We found that convergence for resolutions finer than \texttt{mesh\_delta\_coeff=0.45}.
For our canonical models we adopt a resolution of \texttt{mesh\_delta\_coeff=0.4}.

We control the size of MESA time-steps to prevent the helium nuclear luminosity from changing by more than $0.3$~per-cent between steps, and further require that the timestep not exceed ${\rm d}t \leq \mathrm{yr} (L_0 / L_{\rm He})$, where $L_{\rm He}$ is the helium-burning luminosity and $L_0$ is a parameter we fixed via convergence testing.
This ensures that near the peak of the flash at most $(L_0 \mathrm{yr}) \mathrm{M}_{\rm conv}$ of wave heat is deposited per step.
We performed a series of runs varying $L_0$, and found convergence for $L_0 \leq 3500$. For our canonical models we adopt $L_0 = 3000$.

While not shown here, we have performed additional convergence tests with respect to atmospheric boundary conditions and MESA solver parameters and found neither to matter significantly.
In total our convergence study involved more than three thousand models.

\bibliography{refs}
\bibliographystyle{aasjournal}

\end{document}